\documentstyle[12pt,aasms4]{article}  

\begin{document}

                           % The preamble begins here.
\title{Ultra Deep Survey for Irregular Satellites of Uranus: Limits to Completeness\altaffilmark{1,2} }  
\author{Scott S. Sheppard\altaffilmark{3}, David Jewitt and Jan Kleyna}    
\affil{Institute for Astronomy, University of Hawaii, \\
2680 Woodlawn Drive, Honolulu, HI 96822 \\ sheppard@ifa.hawaii.edu, jewitt@ifa.hawaii.edu, kleyna@ifa.hawaii.edu}

\altaffiltext{1}{Based largely on data collected at Subaru Telescope, which is operated by the National Astronomical Observatory of Japan.}
\altaffiltext{2}{Based in part on observations obtained at the Gemini Observatory, which is operated by the Association of Universities for Research in Astronomy, Inc., under a cooperative agreement with the NSF on behalf of the Gemini partnership.}
\altaffiltext{3}{Current address: Department of Terrestrial Magnetism, Carnegie Institution of Washington, 5241 Broad Branch Rd. NW, Washington, DC 20015; sheppard@dtm.ciw.edu.}

\begin{abstract}  % Produces abstract

We present a deep optical survey of Uranus' Hill sphere for small
satellites.  The Subaru 8-m telescope was used to survey about 3.5
square degrees with a $50 \%$ detection efficiency at limiting red
magnitude $m_R$ = 26.1 mag.  This magnitude corresponds to objects
that are about 7 km in radius (assuming an albedo of 0.04).  We
detected (without prior knowledge of their positions) all previously
known outer satellites and discovered two new irregular satellites
(S/2001 U2 and S/2003 U3).  The two inner satellites Titania and
Oberon were also detected.  One of the newly discovered satellites
(S/2003 U3) is the first known irregular prograde of the planet.  The
population, size distribution and orbital parameters of Uranus'
irregular satellites are remarkably similar to the irregular
satellites of gas giant Jupiter.  Both have shallow size distributions
(power law indices $q \sim 2$ for radii $> 7$ km) with no correlation
between the sizes of the satellites and their orbital parameters.
However, unlike those of Jupiter, Uranus' irregular satellites do not
appear to occupy tight distinct dynamical groups in semi-major axis
versus inclination phase space.  Two groupings in semi-major axis
versus eccentricity phase space appear to be statistically
significant.

\end{abstract}


\section{Introduction}

Planetary satellites are confined to the space in which the planet's
gravitational force dominates over the Sun's.  This region is known as
the Hill sphere, the radius of which, $r_{H}$, is given by

\begin{equation}
r_H = a_p \left[\frac{m_p}{3M_{\odot}}\right]^{1/3}
\label{eq:hill}
\end{equation}

\noindent where $a_p$ and $m_p$ are the semi-major axis and mass of
the planet and $M_{\odot}$ is the mass of the Sun.  Table 1 lists the
Hill sphere radii and projected areas for each of the giant planets.

The giant planets possess two distinct types of satellite (Peale
1999).  Regular satellites are found within about $0.05 r_{H}$ and are
tightly bound to their planet.  They have nearly circular, prograde
orbits with low inclinations.  Regular satellites likely formed within
a circumplanetary disk of gas and dust around the giant planets as
part of the planetary formation process itself.  In contrast,
irregular satellites are found up to $0.65 r_{H}$ from their host
planets and have moderate to high eccentricities and inclinations with
prograde or retrograde orbits.  Irregular satellites can not have
formed in their present orbits and are likely products of early
capture from heliocentric orbit (Kuiper 1956; Pollack, Burns \& Tauber
1979).  Table 1 lists the currently known populations of irregular
satellites for each giant planet as of August 1, 2004.

Burns (1986) offered a definition of irregular satellites as those
satellites which are far enough from their parent planet that the
precession of their orbital plane is primarily controlled by the Sun
instead of the planet's oblateness.  In other words, the satellite's
inclination is fixed relative to the planet's orbit plane instead of
the planet's equator.  By this definition, any satellite with a
semi-major axis larger than the critical value, $a_{\mbox{crit}} \sim
(2 J_{2} r_{p}^{2} a_{p}^{3} m_{p} / M_{\odot} )^{1/5}$, is an
irregular satellite (Burns 1986; see Table 1).  Here $J_{2}$ is the
planet's second gravitational harmonic coefficient and $r_{p}$ is the
planet's equatorial radius.

Because of the reversibility of Newton's equations of motion some sort
of energy dissipation is required for permanent satellite capture.
The giant planets currently have no efficient mechanism of energy
dissipation for satellite capture.  During the planet formation epoch
several mechanisms may have operated to capture satellites: 1) gas
drag in an extended, primordial planetary atmosphere (Pollack, Burns
\& Tauber 1979) 2) pull-down capture caused by the mass growth of the
planet and consequent expansion of the Hill sphere (Heppenheimer \&
Porco 1977) and 3) orbital energy dissipation from collisions or
collisionless interactions between asteroids and/or satellites passing
near the planet (Colombo \& Franklin 1971; Tsui 2000).  Study of the
irregular satellites is important for the insight these objects might
provide into the planet formation process.

Core accretion models of planet formation struggle to form Uranus
and Neptune within the age of the solar system (Lissauer et al. 1995;
Pollack et al. 1996; Boss 2001).  Disk instability models do not
readily provide the hydrogen- and helium- depleted compositions of the
two ice giants (Boss 2001).  The massive hydrogen and helium gas
giants Jupiter and Saturn likely formed quickly in the protoplanetary
disk ($\le 10^{6}$ years).  The less massive, deficient in hydrogen
and helium, more distant ice giants Uranus and Neptune appear to have
taken much longer or an altogether different route of formation
(Thommes et al. 2002; Boss 2002).

Uranus is noteworthy in the sense that its obliquity exceeds 90
degrees.  This compares to the modest obliquities of Jupiter, Saturn
and Neptune at 3, 27 and 30 degrees respectively.  A possible cause is
that a protoplanet of about one Earth mass collided with Uranus near
the end of its growth phase (Korycansky et al. 1990; Slattery et
al. 1992).  Greenberg (1974) argued that the current regular
satellites must have formed after Uranus' obliquity reached 98 degrees
because their low inclination prograde orbits would not have adjusted
to their current configurations with Uranus.  Recently Brunini et
al. (2002) suggested that if Uranus' tilt was created by a giant
impact any satellites beyond about $2 \times 10^{6}$ km (i.e. all
known irregular satellites of Uranus) would have likely been lost
owing to the orbital impulse imparted to Uranus by the impactor.  In
addition, Beauge et al. (2002) show that any significant migration by
Uranus through a residual planetary disk would have caused its outer
satellites to become unstable.

By virtue of its proximity, Jupiter has the best-studied irregular
satellite system (Figure \ref{fig:distance26uranus}), with 55
irregular satellites currently known (Sheppard \& Jewitt 2003).  In
this paper we ask the question ``does the ice-giant Uranus have a
population of irregular satellites similar to that of gas-giant
Jupiter?''.  The greater distance of Uranus requires the use of very
deep surveys in order to meaningfully probe the smaller Uranian
satellites.  Previous surveys near Uranus were conducted using 4-meter
class telescopes, including a search of $\sim$5 deg$^2$ to limiting
red magnitude $m_R \sim$ 24.3 by Gladman et al. (2000) and of $\sim$1
deg$^2$ to limiting magnitudes in the $m_R \sim$ 25.0 to $m_R \sim$
25.4 range by Kavelaars et al. (2004).  In the present work, we used
the 8-meter Subaru telescope and its prime-focus survey camera to
survey most of the Hill sphere to a limiting red magnitude $m_R$ =
26.1.  Our primary goal was to cover the dynamically stable inner Hill
sphere of the planet (radial extent $\sim 0.7 r_{H}$; see Hamilton \&
Krivov 1997) in an unbiassed, deep and uniform survey.

\section{Observations}

We observed the space around Uranus for faint satellites near new moon
on UT August 29 and 30, 2003 with the Subaru 8.2 meter diameter
telescope atop Mauna Kea.  The geometry of Uranus during the survey is
indicated in Table~2.  The Suprime-Cam imager has 10 MIT/LL $2048
\times 4096$ CCDs arranged in a $5 \times 2$ pattern (Miyazaki et
al. 2002).  Its $15 \micron$ pixels give a scale of $0.\arcsec 20$
pixel$^{-1}$ at prime focus and a field-of-view that is about
$34\arcmin \times 27\arcmin$ with the North-South direction aligned
with the long axis.  Gaps between the chips are about $16 \arcsec$ in
the North-South direction and only $3 \arcsec$ in the East-West
direction.

Images were obtained with a Kron-Cousins R-band filter (central
wavelength near 650 nm).  The images were bias subtracted and then
flat-fielded with dome flats taken at the end of each night.  During
exposures the telescope was autoguided sidereally on field stars.
Seeing during the two nights varied between $0.\arcsec 4$ and
$0. \arcsec 6$ Full Width at Half Maximum (FWHM).  Integration times
were between 400 and 420 seconds.  Both nights were photometric and
Landolt (1992) standards were used for calibration.

The area searched around Uranus for satellites is shown in
Figure~\ref{fig:areauranus}.  Fourteen fields were imaged 3 times each
on one night and 2 times each on the second night for a total of 5
images per field or 70 images for the survey.  On a given night,
images of the same field were separated in time by about 31 minutes.
The fields on the second night were centered at the same angular
distance from Uranus as on the first night, but the background star
fields were different because of Uranus' non-sidereal motion.
Approximately 3.5 square degrees around Uranus were observed, not
accounting for chip gaps and bright stars.  We covered $\sim 90 \%$ of
the theoretically stable Hill sphere ($\sim 0.7 r_{H}$) for retrograde
satellites and near $100 \%$ for the stable Hill sphere ($\sim 0.5
r_{H}$) for prograde satellites (Hamilton \& Krivov 1997).

Satellite confirmation astrometry in the months after discovery was
obtained with the Gemini Multi-Object Spectrograph (GMOS) in imaging
mode on the Gemini North 8.1 meter telescope atop Mauna Kea (Hook et
al. 2002).  GMOS has a $5.\arcmin 5$ field of view and pixel scale of
$0.\arcsec 0727$ pixel$^{-1}$.  Recoveries were obtained with an r'
filter based on the Sloan Digital Sky Survey (SDSS) filter set.
Images were bias subtracted and flat-fielded using dome flats.
Observations were obtained by a staff astronomer at Gemini in queue
scheduling mode.

\section{Analysis}

The Subaru survey observations were obtained when Uranus was near
opposition, where the apparent movement is largely parallactic and
thus is inversely related to the distance.  Objects at the
heliocentric distance of Uranus, $R\sim 20$ AU, will have an apparent
motion of about $\sim 6\arcsec $~hr$^{-1}$ ($\sim 30$ pixels per
hour).  At this rate satellites would have trailed a distance
comparable to the FWHM of the seeing during the 400 second exposures.

The data were analyzed to find solar system bodies in two
complementary ways.  First a computer algorithm was used to detect
objects which appeared in all three images from one night and which
had a motion consistent with being beyond the orbit of Jupiter (speeds
less than 20 arcseconds per hour).  Second, all the fields were
examined by visually blinking them on a computer display screen for
moving objects again with motions indicative of distances beyond
Jupiter.

We determined the limiting magnitude of the survey in the absence of
scattered light from Uranus by placing artificial objects in the
fields matched to the point spread function of the images and with
motions mimicking that of Uranus.  The brightnesses of the objects
were binned by 0.1 mag and spanned the range from 25 to 27 magnitudes.
Results are shown in Figure \ref{fig:effuranus} for both the visual
blinking and computer algorithm.  The techniques gave similar results,
though the visual blinking was slightly more efficient in detection.
The $50 \%$ differential detection efficiency of the Uranus satellite
survey occurred at an R-band limiting magnitude of about 26.1
magnitudes which we take as the limiting magnitude of this survey.
Scattered light from Uranus was only significant within 3.5 arcminutes
of the planet.  Figure \ref{fig:effuranusbright} shows the $50 \%$
R-band limiting magnitude efficiency as a function of distance from
Uranus in our survey.  The scattered light results were determined in
the same way as the nonscattered light technique described above.

\section{Results and Discussion}

Our survey detected, without prior knowledge of their positions, all
of the then-known six irregular satellites of Uranus as well as the
two outer-most regular satellites Titania and Oberon.  We also
discovered the two new Uranian satellites S/2001 U2 and S/2003 U3
(Sheppard et al. 2003).  S/2001 U2 has a 2001 designation because it
had been detected but not confirmed as a Uranian satellite in 2001
(Holman et al. 2003).  S/2001 U3 was confirmed as a Uranian irregular
satellite after our survey (Marsden et al. 2003).  This object went
undetected in our Subaru data because it fell on a bright star during
observations.  In addition, six other objects near Uranus which turned
out to be Centaurs, as well as hundreds of Kuiper Belt objects, were
discovered.  A detailed report about the discovery of these other
objects will be given in a future paper.

To determine the size limit of satellites detectable by the survey and
the approximate sizes of the new satellites we relate the apparent red
magnitude, $m_{R}$, to the radius, $r$, through

\begin{equation}
r = \left[ \frac{2.25\times 10^{16}R^{2}\Delta ^{2}}{p_{R}\phi
(\alpha)} \right]^{1/2} 10^{0.2(m_{\odot} - m_{R})}
\label{eq:appmaguranus}
\end{equation}

\noindent in which $r$ is in km, $R$ is the heliocentric distance in
AU, $\Delta$ is the geocentric distance in AU, $m_{\odot}$ is the
apparent red magnitude of the sun ($-27.1$), $p_{R}$ is the geometric
red albedo, $\phi (\alpha)$ is the phase function and $\alpha$ is the
phase angle ($\alpha=0$ deg at opposition).  For linear phase
functions we use the notation $\phi (\alpha) = 10^{-0.4 \beta
\alpha}$, where $\beta$ is the ``linear'' phase coefficient.  Using
data from Table~2 and an albedo of 0.04 we find that 26.1 magnitudes
corresponds to a satellite with radius of about 7 km.

\subsection{Size and Population Distribution}

The Cumulative Luminosity Function (CLF) describes the sky-plane
number density of objects brighter than a given magnitude.  The CLF is
conveniently described by

\begin{equation}
\mbox{log}[\Sigma (m_{R})]=\alpha (m_{R}-m_{o})  \label{eq:slope}
\end{equation}

\noindent where $\Sigma (m_{R})$ is the number of objects brighter
than $m_{R}$, $m_{o}$ is the magnitude zero point, and $\alpha$
describes the slope of the luminosity function.  The CLF for Uranus'
irregular satellites is shown in Figure~\ref{fig:cumuran}.  Our survey
is complete to $m_R$ = 26.1 mag ($r > 7$ km).  For Uranus we find the
best fit to the CLF for $m_{R} < 26$ mag is $\alpha = 0.20 \pm 0.04$
and $m_{o} = 20.73 \pm 0.2$ (Figure \ref{fig:cumuran}).

The points in a CLF are heavily correlated with one another, tending
to give excess weight to the faint end of the distribution.  The
Differential Luminosity Function (DLF) does not suffer from this
problem.  We plot the DLF using a bin size of 2 mag for all Uranian
irregular satellites in Figure \ref{fig:diffuran}.  We find that the
DLF has a slope of $\alpha = 0.16 \pm 0.05$ with $m_{o} = 21.0 \pm
0.4$.  Bin sizes of 1 mag and 1.5 mag give similar results.  The
uncertainty of the fit for the 2 mag bin size is estimated from the
fits of these other bin sizes which were larger than the uncertainty
of the least square fit for the 2 mag bin size alone.

The luminosity functions represent the combined effects of the
albedos, heliocentric distances and size distributions of the objects.
If we assume that all Uranian satellites are at the same heliocentric
distance and that their albedos are similar, the DLF and CLF simply
reflect the size distribution of the satellites.  The objects appear
to follow a single power-law size distribution for $r > 7$ km.  In
order to model the irregular satellite size distribution we use a
differential power-law radius distribution of the form $n(r)dr=\Gamma
r^{-q}dr$, where $\Gamma$ and $q$ are constants, $r$ is the radius of
the satellite, and $n(r)dr$ is the number of satellites with radii in
the range $r$ to $r+dr$.  The slope of the DLF ($\alpha$) and exponent
of the size distribution ($q$) are simply related as $q = 5 \alpha +
1$ when assuming similar heliocentric distance and albedos for all
satellites (Irwin et al. 1995).  Using $\alpha = 0.16 \pm 0.05$ for
Uranus' irregular satellites we find $q = 1.8 \pm 0.3$.  This is
similar to the value found by Kavelaars et al. (2004) and identical to
$q = 1.9 \pm 0.3$ found for the irregular satellites of Jupiter in the
same size range (Sheppard \& Jewitt 2003).

In comparison, collisional equilibrium gives $q \sim 3.5$ (Dohnanyi
1969), nonfamily small asteroids have $q \sim 2.0$ to $2.5$ (Cellino
et al. 1991), while large Kuiper Belt Objects (KBOs) have $q=4.2\pm
0.5$, $\alpha = 0.64 \pm 0.1$, $m_{o} = 23.23 \pm 0.15$ (Trujillo et
al. 2001) and Centaurs have KBO-like slope with $m_{o} = 24.6 \pm 0.3$
(Sheppard et al. 2000).  The small Jovian Trojans ($r<30$ km) have
$q=3.0 \pm 0.3$ while the larger Trojans show a steeper slope of
$q=5.5 \pm 0.9$ (Jewitt, Trujillo \& Luu 2000).  Large members of
asteroid families have been found to usually have $q \ge 4$ (Tanga et
al. 1999) while the smaller members ($r < 5$ km) may have shallower
distributions (Morbidelli et al. 2003).  If $q>3$, most of the
collisional cross-section lies in the smallest objects while for $q>4$
most of the mass lies in small bodies.  The mass and cross-section of
Uranus' irregular satellites are both dominated by the few largest
objects.

\subsection{Comparison of the Uranian and Jovian Systems}

The Jovian and Uranus satellite DLFs are compared in Figure
\ref{fig:diffuran}.  Fitting the Jovian DLF (which is complete to
around $r \sim 3$ km; Sheppard \& Jewitt 2003) over the same size
range as that used previously for Uranus ($r > 7$ km) we find a slope
of $\alpha = 0.18 \pm 0.05$ with a zero point magnitude $m_{o} = 14.0
\pm 0.4$.  The measured slope is compatible with $\alpha = 0.16 \pm
0.05$, as found for the Uranian satellites, within the statistical
uncertainties.  We conclude that the size distribution indices of the
irregular satellites at Uranus and Jupiter are remarkably similar for
$r > 7$ km, and both are quite different from the (steeper)
distribution that would be measured, for example, amongst the
main-belt asteroids.  Jupiter's irregular satellites appear depleted
in the 4 to 10 km size range relative to an extrapolation of a
power-law fitted at larger sizes (Sheppard and Jewitt 2003).  We would
need deeper survey observations to $m_R \sim$ 27.2 mag to determine
whether the Uranian population shows a similar depletion.

The difference between the Jovian and Uranian DLF magnitude zero
points is $\Delta m_0$ = 7.0$\pm$0.6 mag.  This is to be compared with
the magnitude difference expected from the inverse square law and the
different distances to the two planets.  The expected magnitude
difference is $\Delta m$ = 5 $log_{10}\left[\frac{R_U(R_U -
1)}{R_J(R_J - 1)}\right]$, where $R_J$ = 5.3 AU and $R_U$ = 20.0 AU,
are the heliocentric distances to the two planets and opposition
geometry is assumed.  Substituting, we obtain $\Delta m$ = 6.1 mag.,
which is different from $\Delta m_0$ by only about 1.5$\sigma$.  In
this sense, the smaller number of known irregular satellites at Uranus
seems to be an artifact of the greater distance.

To emphasize these points, we show the cumulative size distributions
of the Jovian and Uranian irregular satellites in Figure
\ref{fig:cumuransize}. Uranus has nine satellites with r$ \geq$ 7 km
(about the completeness level) while Jupiter has eight.  In terms of
the size distributions and total populations, the irregular satellite
systems of Uranus and Jupiter are very similar.  If we assume the size
distribution of Uranus' irregular satellites extends down to radii of
about 1 km, we would expect about $75 \pm 30$ irregular satellites of
this size or larger.

Several competing processes could influence the size distribution of
the satellites.  The larger objects may retain some memory of the
production function, as is apparently the case with the nearby Jovian
Trojans, in which the size distribution is steeper below $r \sim$ 30
km than above it (Jewitt et al. 2000).  Small satellites could be lost
to gas drag, leading to a flattening of the size distribution.  This
is more likely at gas-giant Jupiter than at ice-giant Uranus, where
much less gas is thought to have been available during the accretion
epoch.  Collisions between satellites and with interplanetary
projectiles would lead to the production of many small fragments.
Given that fragment velocity and size are inversely related, it is
natural to expect that the smaller objects produced collisionally
would be lost (the escape velocity from Uranus at the semi-major axis
of S/2001 U2 is only 0.8 km s$^{-1}$), again leading to a flattened
size distribution.  Collisional scenarios, in general, require higher
collision rates than now prevail in the solar system.  Perhaps the
irregular satellites were originally much more numerous than now.
Separately, we know that the flux of planet-crossing projectiles was
much higher between the epochs of planet formation and the end of the
terminal bombardment phase at about 3.9 Gyr.

\subsection{Orbital Element Distribution}

Table 3 lists some of the properties of the known irregular satellites
of Uranus.  Figures \ref{fig:irrsatsuranusecc} and
\ref{fig:irrsatsuranus} compare the semi-major axes with inclinations
and eccentricities, respectively, for all known irregular satellites
of the planets.  The figures show that the ice giants Uranus and
Neptune have the smallest known irregular satellite systems, in units
of Hill radii.  In the case of Uranus, one contributing factor may be
that its regular satellite system is much less massive and does not
extend as far from the planet as those of the gas giant planets of
Jupiter and Saturn.  Thus, interactions between Uranus' regular and
irregular satellites are less important as a clearing mechanism.  We
also note that Uranus has the smallest $a_{\mbox{crit}}$ of any of the
giant planets (Table 1).  Neptune's satellite system may have been
severely disrupted by the unusually large retrograde satellite Triton
(Goldreich et al. 1989).

Figure \ref{fig:irrsatsuranusecc} suggests that the Uranian irregular
satellites may be grouped in semi-major axis vs. eccentricity phase
space.  The four retrograde irregular satellites closest to Uranus
(S/2001 U3, Caliban, Stephano and Trinculo; semi-major axes of $a <
0.15 r_{H}$) have eccentricities $e \sim$0.2 while the four retrograde
irregular satellites (Sycorax, Prospero, Setebos and S/2001 U2) with
$a > 0.15 r_{H}$ have $e \sim$0.5.

To judge the significance of these two retrograde groups we performed
several statistical tests.  The retrograde low eccentricity, low
semi-major axis group (the Caliban group) has a mean eccentricity of
$0.19 \pm 0.02$ and mean semi-major axis of $7.0\pm 0.9 \times 10^{6}$
km while the retrograde high semi-major axis and eccentricity group
(the Sycorax group) has $0.52 \pm 0.05$ and $16.7 \pm 1.8 \times
10^{6}$ km, respectively.  The Student's t-test (with 7 degrees of
freedom) gave a t-statistic of 6.0 with a significance of 99.8\% for
the difference in their mean eccentricities and a t-statistic of 4.8
with a significance of 99.4\% for the difference in their mean
semi-major axes.  The student's t-test suggests that the two groups
are significant but makes the unjustified assumption that the
eccentricity and semi-major axes are normally distributed.  Therefore,
we used the more stringent nonparametric Mann-Whitney U-test and a
permutation test (see Siegel \& Castellan 1988) to assess the
significance of the two groups.  Both found that the groupings of the
eccentricity were statistically significant at or above the $\geq
3\sigma$ (99.7\%) level of confidence.

Figure \ref{fig:ecccorruran} shows the two retrograde groups in
semi-major axis vs. eccentricity space.  It is clearly seen that the
more distant satellites have larger eccentricities as was also noted
by Kavelaars et al. (2004).  The significance of a linear fit to the
data is only at the $\sim$97 \% ($< 3 \sigma$) level and thus less
significant than the two groupings.  Possible reasons for higher
eccentricities more distant from the planet is that the closer
satellites would be unstable to perturbations by the much larger
regular satellites of Uranus if they had large eccentricities or that
the more distant satellites are susceptible to solar and planetary
perturbations.  In Figure \ref{fig:ecccorruran} we plot several lines
of constant periapse from Uranus.

The orbital velocities of the irregular satellites around Uranus range
from 300 to 1100 m/s (Kessler 1981).  The relative velocities amongst
the satellites are typically much smaller.  Velocity differences among
Caliban, Stephano and Trinculo  are comparable to Caliban's escape
velocity $\sim$40 m/s.  The retrograde high eccentricity objects have
relative velocities comparable to Sycorax's escape velocity of
$\sim$80 m/s.  For comparison, Jupiter's irregular satellites were
found to be grouped in semi-major axis and inclination phase space
with their relative velocities within a group about 30 m/s while group
velocities relative to each other were over 200 m/s (Sheppard \&
Jewitt 2003).  As at Jupiter, Uranus' possible two retrograde groups
have relative velocities of over 100 m/s while members within a group
have velocities comparable to the largest members escape velocity.  As
at Jupiter, the dynamical groupings suggest formation from parent
objects which were collisionally shattered.  A simple particle in a
box calculation shows that collisions between the currently known or
predicted outer satellite population of Uranus would occur on
timescales ($\sim 10^{10}$ years) longer than the age of the solar
system.  Fragmentation could have occurred from collisions with
objects in heliocentric orbits (principally comets around the heavy
bombardment period (Sheppard \& Jewitt 2003)) or other now defunct
satellites of the planet (Nesvorny et al. 2004).  Jupiter's irregular
satellite orbital velocities ($> 2200$ m/s) are much greater than
Uranus' and thus any collisional processing would have been much more
violent compared to collisions around Uranus.

Except for the distinct prograde irregular S/2003 U3 with an orbital
inclination of 57 degrees to the ecliptic compared to the eight
irregulars between 140 and 170 degrees there are no obvious tight
groupings in semi-major axis vs. inclination phase space as was found
around Jupiter (Figure \ref{fig:irrsatsuranus}).  We do note that the
majority of the low eccentricity retrograde satellites have
inclinations near 142 degrees while the retrograde high eccentricity
satellites are closer to 156 degrees in inclination (Table 3).  The
intermediate inclinations $60 < i < 140$ degrees are devoid of known
satellites, consistent with instabilities caused by the Kozai
instability (Kozai 1962; Carruba et al. 2002; Nesvorny et al. 2003).
In this instability, solar perturbations at apoapse cause the
satellites at high inclinations to acquire large eccentricities which
eventually lead to collisions with the planet, or a regular satellite
or loss from the Hill sphere in $10^{7} - 10^{9}$ years (Carruba et
al. 2002; Nesvorny et al. 2003).

We find no clear size vs. semi-major axis, inclination or eccentricity
correlations for Uranus' irregular satellites, as may be expected if
significant gas drag was present in the past.  Of the two largest
irregular satellites around Uranus one of each is in the two possible
eccentricity groups.  Caliban is relatively close to the planet with a
low eccentricity while Sycorax is with the distant, higher
eccentricity irregular satellites (Table 3).

At Uranus there are many more known retrograde (8) than prograde (1)
outer satellites.  Jupiter also has an over abundance of known
retrograde outer satellites (48 retrograde versus 7 prograde).  These
asymmetries are greatly diminished if we compare numbers of satellite
groups ($\sim 3-4$ retrograde versus 3 prograde groups at Jupiter and
possibly 1 or 2 retrograde and 1 prograde group at Uranus).  Given
this, and the fact that the statistics of the groups remain poor
(especially at Uranus), we cannot currently use the relative numbers
of retrograde and prograde objects to constrain the mode of capture.

\section{Summary}

1) We have conducted a deep imaging survey of 3.5 deg$^2$ around
Uranus covering most of the region in which long-term stable orbits
are possible.  The effective limiting red magnitude of the survey is
$m_R$ = 26.1 mag (50\% detection efficiency).  This corresponds to
objects of about 7 km in radius if assuming a 0.04 geometric albedo.

2) We detected, without prior knowledge of their positions, all
previously known irregular satellites in addition to two new irregular
satellites (S/2001 U2 and S/2003 U3). The latter is Uranus' first and,
so far, only known prograde irregular satellite.

3) The differential size distribution of the irregular satellites
approximates a power law with an exponent $q$ = 1.8$\pm$0.3 (radii $>
7$ km).  In this relatively flat distribution, the cross-section and
mass are dominated by the few largest members.  The size distribution
is essentially the same as found independently for the irregular
satellites of the gas giant Jupiter.

4) The Jovian and Uranian irregular satellite populations, when
compared to a given limiting size, are similar.  For example, the
number of satellites larger than 7 km in radius (albedo 0.04 assumed)
is nine at Uranus compared with eight at Jupiter.  The similarity is
remarkable given the different formation scenarios envisioned for
these two planets.

5) The orbital parameters of the satellites are unrelated to their
   sizes.

6) We tentatively define two groups of retrograde irregular satellites
in semi-major axis vs. eccentricity phase space. The four satellites
of the inner retrograde group (Caliban, S/2001 U3, Stephano and
Trinculo) have semi-major axes $<$0.15 Hill radii and moderate
eccentricities ($\sim$0.2) while the four members of the outer
retrograde group (Sycorax, Prospero, Setebos and S/2001 U2) have
larger semi-major axes ($>$ 0.15 Hill radii) with higher
eccentricities ($\sim$0.5).  Unlike at Jupiter, the currently known
retrograde irregular satellites are not tightly grouped in semi-major
axis versus inclination space.

\section*{Acknowledgments}

We thank Richard Wainscoat for confirming coordinates for Gemini and
the Gemini staff for running the queue servicing mode with GMOS.  We
are grateful to Brian Marsden and Bob Jacobson for orbital
calculations relating to the satellites and to the anonymous referee
for a careful review.  This work was supported by a grant from NASA to
DJ.

\newpage

\begin{figure}
\caption{The area searched around Uranus for satellites using the 8.2m
Subaru telescope.  Fourteen fields were imaged on five occasions over
two nights (UT August 29 and 30, 2003) for a total of 70 images
covering about 3.5 square degrees.  The black dot at the center
represents Uranus' position.  Uranus was placed in a gap in the mosaic
of CCD chips to prevent saturation.  The dotted circle shows the Hill
sphere of Uranus while the dashed circle shows the theoretical outer
limits of long-term stability for any Uranus satellites ($r_{H} \sim
0.7$).  The outer satellites of Uranus are marked at the position of
their detection during the survey.  In addition we detected the inner
satellites Titania and Oberon (not marked).}
\label{fig:areauranus} 
\end{figure}

\begin{figure}
\caption{The distances of the planets versus the observable small body
population diameter for a given red magnitude assuming a low (0.04)
geometric albedo.  The mean semi-major axes of the giant planets
Jupiter (J), Saturn (S), Uranus (U) and Neptune (N) are marked.}
\label{fig:distance26uranus} 
\end{figure}

\begin{figure}
\caption{Detection efficiency of the Uranus survey versus the apparent
red magnitude.  The 50\% detection efficiency is at about 26.1 mag
from visual blinking and 26.0 mag from a computer program.  All fields
were searched with both techniques.  The efficiency was determined by
placing artificial objects matched to the Point Spread Function (PSF)
of the images with motions similar to Uranus in the survey fields.
Effective radii were calculated assuming the object had an albedo of
0.04.  The efficiency does not account for objects which would have
been undetected because of the chip gaps.}
\label{fig:effuranus} 
\end{figure}

\begin{figure}
\caption{The red limiting magnitude (50\% detection efficiency) of the
survey versus distance from Uranus.  Scattered light is only
significant starting at about 3.5 arcminutes from Uranus. The
calculation of the effective radius assumes an albedo of 0.04.  }
\label{fig:effuranusbright} 
\end{figure}

\begin{figure}
\caption{The cumulative luminosity function (CLF) of the irregular
satellites of Uranus.  The dotted line is the slope of the CLF for
Uranus' irregular satellites with $m_{R} \le 26$ mag ($\alpha = 0.20
\pm 0.04$).  }
\label{fig:cumuran} 
\end{figure}

\begin{figure}
\caption{The differential luminosity function (DLF) of the irregular
satellites of Uranus and Jupiter.  The slope for both planets is very
similar but because of Uranus' further distance it is shifted about 6
magnitudes to the right.  The dotted line is the slope of the DLF for
Uranus' irregular satellites with $r \ge 7$ km ($\alpha = 0.16 \pm
0.05$) and the dashed line is for Jupiter's irregular satellites with
the same size range ($\alpha = 0.18 \pm 0.05$).  A deficiency of
satellites around Jupiter with magnitudes in the range $19 < m_{R} <
21.5$ ($ 4 < r < 10$ km) is clearly seen as described in Sheppard \&
Jewitt (2003).}
\label{fig:diffuran} 
\end{figure}

\begin{figure}
\caption{The cumulative radius function for the irregular satellites
of Uranus and Jupiter.  This figure directly compares the sizes of the
satellites of the two planets assuming both satellite populations have
albedos of about 0.04.  The two planets have statistically similar
size distributions of irregular satellites ($q \sim 2$) for a size
range of $r \ge 7$ km.}
\label{fig:cumuransize} 
\end{figure}

\begin{figure}
\caption{A mean eccentricity comparison between the known irregular
satellites of the giant planets.  The horizontal axis is the ratio of
the satellite's mean semi-major axis to its respective planet's Hill
radius.  The vertical axis is the mean eccentricity of the satellite.
All giant planets independent of their mass or formation appear to
have similar irregular satellite systems.  Except for Neptune's
``unusual'' Triton and Nereid, Uranus appears to have the closest
irregular satellites in terms of Hill radii of the giant planets.  }
\label{fig:irrsatsuranusecc} 
\end{figure}

\begin{figure}
\caption{ Same as Figure \ref{fig:irrsatsuranusecc} except here mean
inclination to the ecliptic is used instead of eccentricity for the
vertical axis.}
\label{fig:irrsatsuranus} 
\end{figure}

\begin{figure}
\caption{The orbital eccentricity versus the mean semi-major axis in
units of the Hill radius for retrograde Uranus irregulars only.  There
is a trend for satellites to have a larger eccentricity if they are
more distant from Uranus.  Boxes show and are named after the largest
member of the two possible groupings discussed in the text.  A simple
linear fit gives a slope of $2.0\pm 0.7$ and a y-intercept at $0.0\pm
0.1$ but the Pearson correlation coefficient is only 0.76,
corresponding to a statistical significance $<$3$\sigma$.  The two
groupings are statistically more significant than the linear fit.
Dashed lines show constant periapse distances of 0.02, 0.05, 0.10,
0.15 and 0.20 Hill radii, respectively.  The regular satellites of
Uranus are found inside 0.02 Hill radii.}
\label{fig:ecccorruran} 
\end{figure}

\end{document}